\documentclass[]{aa}

\usepackage{ulem}
\usepackage{color}
\usepackage{graphics,graphicx}
\usepackage{times} 
\usepackage{amssymb}
\usepackage{amsmath}
\usepackage{natbib}
\usepackage{lscape}
\usepackage{url}
\usepackage{multirow}
\usepackage{ctable}
\usepackage{threeparttable}
\usepackage[labelfont=bf,labelsep=period]{caption}

\usepackage{xspace}

\def\xmm{\textsl{XMM-Newton}\xspace}

\def\chandra{\textsl{Chandra}\xspace}
\def\swift{\textsl{Swift}\xspace}
\def\swiftng{{\it Neil Gehrels Swift Observatory}}

\def\epicpn{{EPIC-pn}}
\def\epicmos1{{EPIC-MOS1}}
\def\epicmos2{{EPIC-MOS2}}
\def\epicmos{{EPIC-MOS}}

\def\nustar{\textsl{NuSTAR}\xspace}

\def\rsph{\ensuremath{R_\text{sph}}\xspace}
\def\mdotin{\ensuremath{\dot M_\text{in}}\xspace}
\def\mdottot{\ensuremath{\dot M_\text{tot}}\xspace}
\def\mdotincrit{\ensuremath{\dot M_\text{in,crit}}\xspace}
\def\mdot{\ensuremath{\dot{M}}\xspace}

\def\ergcms{\ensuremath{\text{erg\,cm}^{-2}\,\text{s}^{-1}}\xspace}
\def\ergs{\ensuremath{\text{erg\,s}^{-1}\xspace}}
\def\msun{\ensuremath{\rm M_{\odot}}\xspace}
\def\ledd{\ensuremath{L_\text{Edd}}\xspace}
\def\mdotedd{\ensuremath{\dot M_\text{Edd}}\xspace}

\def\barycen{\hbox{\sc barycen}}
\def\sas{\hbox{\rm{\small SAS~\/}}}
\def\epchain{\hbox{\sc epchain}}
\def\xmmselect{\hbox{\sc xmmselect}}

\def\apj{ApJ}
\def\mnras{MNRAS}
\def\nat{Nat}

\def\araa{ARA\&A}                
\def\aap{A\&A}                   
\def\aj{AJ}                      
\def\apjl{ApJ}                   

\def\ngc{NGC\,5907 ULX1\xspace}

\begin{document}

\title{Probing the nature of the low state in the extreme ultraluminous X-ray pulsar \ngc}

\author{F.~F\"urst\inst{1}\and D.~J.~Walton\inst{2,3} \and G.~L.~Israel\inst{4} \and 
M.~Bachetti\inst{5} \and
D.~Barret\inst{6} \and
M.~Brightman\inst{7} \and
H.~P.~Earnshaw\inst{7} \and
A.~Fabian\inst{3} \and 
M.~Heida\inst{8} \and 
M.~Imbrogno\inst{9,4}
M.~J.~Middleton\inst{10} \and
C.~Pinto\inst{11} \and
R.~Salvaterra\inst{12} \and
T.~P.~Roberts\inst{13} \and 
G.~A.~Rodríguez~Castillo\inst{11} \and
N.~Webb\inst{6} 
}

\institute{Quasar Science Resources SL for ESA, European Space Astronomy Centre (ESAC), Science Operations Departement, 28692 Villanueva de la Ca\~nada, Madrid, Spain
\and Centre for Astrophysics Research, University of Hertfordshire, College Lane, Hatfield AL10 9AB, UK
\and Institute of Astronomy, Madingley Road, Cambridge CB3 0HA, UK
\and INAF -- Osservatorio Astronomico di Roma, via Frascati 33, 00078 Monteporzio Catone, Italy
\and INAF - Osservatorio Astronomico di Cagliari, via della Scienza 5, 09047 Selargius, Italy
\and CNRS, IRAP, 9 Av. colonel Roche, BP 44346, 31028 Toulouse cedex 4, France
\and Cahill Center for Astronomy and Astrophysics, California Institute of Technology, Pasadena, CA 91125, USA
\and European Southern Observatory, Karl-Schwarzschild-Strasse 2, 85748 Garching bei M\"unchen, Germany
\and Department of Physics, University of Rome ``Tor Vergata,'' Via della Ricerca Scientifica 1, 00133 Rome, Italy
\and Department of Physics and Astronomy, University of Southampton, Highfield, Southampton SO17 1BJ, UK
\and INAF - IASF Palermo, Via U. La Malfa 153, 90146 Palermo, Italy
\and INAF - IASF Milano, Via Corti 12, 20133 Milano, Italy
\and Centre for Extragalactic Astronomy \& Department of Physics, Durham University, South Road, Durham DH1 3LE, UK}

\abstract{
\ngc is the most luminous ultra-luminous X-ray pulsar (ULXP) known to date, reaching luminosities in excess of $10^{41}$\,\ergs. The pulsar is known for its fast spin-up during the on-state. Here, we present a long-term monitoring of the X-ray flux and the pulse period between 2003--2022. We find that the source was in an off- or low-state between mid-2017 to mid-2020. During this state, our pulse period monitoring shows that the source had spun down considerably. We interpret this spin-down as likely being due to the propeller effect, whereby accretion onto the neutron star surface is inhibited. 
Using state-of-the-art accretion and torque models, we use the spin-up and spin-down episodes to constrain the magnetic field. For the spin-up episode, we find solutions for magnetic field strengths of either around $10^{12}$\,G or $10^{13}$\,G, however, the strong spin-down during the off-state seems only to be consistent with a very high magnetic field, namely, $>10^{13}$\,G. This is the first time a strong spin-down is seen during a low flux state in a ULXP. Based on the assumption that the source entered the propeller regime, this gives us the best estimate so far for the magnetic field of \ngc.
}


\date{Received XX.XX.XX / Accepted XX.XX.XX}

\maketitle

\section{Introduction}

A growing number of ultraluminous X-ray sources (ULXs), namely, off-nuclear X-ray binaries
with apparent luminosities that exceed $10^{39}$\,\ergs (see \citealt{Kaaret17rev}
for a review), are now known to be powered by highly magnetized neutron star
accretors (with magnetic fields of $B\gg10^9$\,G). Six such sources have been discovered to date through the detection of
coherent X-ray pulsations and  thus classified as ultra-luminous X-ray pulsars (ULXPs): M82 X-2 (\citealt{Bachetti14nat}), NGC\,7793 P13
(\citealt{Fuerst16p13, Israel17p13}), \ngc (\citealt{Israel17}), NGC\,300
ULX1 (\citealt{Carpano18}), NGC\,1313 X-2 (\citealt{Sathyaprakash19}) and M51 ULX7
(\citealt{Rodriguez20}). A few more candidates have been found through the tentative identification of pulsations  \citep[e.g., NGC\,7793 ULX-4,][]{quintin21a} or the possible detection of cyclotron resonant scattering lines \citep[e.g., M51 ULX-8,][]{Brightman18}.
Additionally, the galactic source Swift~J0243.6+6124 reached luminosities significantly above  $10^{39}$\,\ergs and can therefore also be classified as a ULXP \citep{wilsonhodge18a, tsygankov18a}. Other similar sources include transient neutron stars with Be-star mass donors in the SMC, like SMC~X-3 \citep{townsend17a, liu22a} and RX~J0209.6$-$7427 \citep{Vasilopoulos20a, Hou22a}, which also reached ULX level luminosities briefly during giant outbursts. While it is currently unclear if the accretion geometry during these outbursts is similar to persistent ULXPs or not (e.g., all ULXPs currently exhibit much higher luminosities), it shows that there is a clear connection between Be-X-ray binaries and ULXPs.

There is speculation that neutron stars could make
up a significant fraction of the total ULX population (e.g., \citealt{Pintore17,
Koliopanos17, King17ulx, Middleton17, Walton18ulxBB}). Furthermore, there is significant
debate over how exactly these neutron stars are able to reach such extreme
luminosities and most of this debate is focused on the strength of the magnetic fields in
these systems. Some authors argue for extremely strong, magnetar-level fields ($B>10^{13}$\,G; e.g., \citealt{Eksi15, Mushtukov15}), as such fields suppress the electron scattering
cross-section (\citealt{Herold79}) and permit higher luminosities for a given
accretion rate, while others argue for much lower fields and accretion rates
($B\approx10^{10}$--$10^{11}$\,G; e.g., \citealt{Kluzniak15, King17ulx}) based on the large spin-up rates ($\dot{P}$)
observed. 

Both scenarios, however, present some issues. The very strong magnetic field scenarios struggle to explain how accretion at luminosities below $\sim10^{40}\,\ergs$ is possible since (in theory) the magnetosphere would stop accretion and the source would be in a permanent propeller state (see below). The low-magnetic field case on the other hand typically also assumes a very high beaming factor to explain the apparent extreme luminosities. This beaming would result in a very narrow funnel, which appears to be in contradiction to the observed sinusoidal pulse profiles with high pulsed fractions \citep{musthukov21a}. Additionally, the very high spin-up rate requires an intrinsically high accretion rate, setting up an argument against an extreme beaming factor.

It is also possible that a combination of these two explanations is present, with a very strong quadrupolar field acting close to the neutron star, whereas further away, the weaker dipolar field dominates \citep[e.g.,][]{Israel17, Middleton19a,Kong22a}.
This scenario has been claimed for Swift~J0243.6+6124, driven by the discovery of a cyclotron resonant scattering feature (CRSF) at up to 146\,keV \citep{Kong22a}. This line energy implies a filed strength in the line-forming region (which is likely to be very close to the neutron star surface) of around $1.6\times10^{13}$\,G. \citet{Kong22a} have argued that such a strong field has to be in the multipolar component, as a dipole of this magnitude would lead to contradictions with other measurements.  Nonetheless, it is currently unclear if the same scenario can explain  the persistent ULXPs at even higher luminosities.

Reliably determining the magnetic fields of ULXPs has proven challenging to date.
In general, the most robust method for determining field strengths in accreting neutron
stars is via the study of cyclotron resonant scattering features (CRSFs; see
\citealt{staubert19a} for a recent review). However, such features are challenging
to detect in ULXs given their relatively low fluxes and the limited bandpass available
for sensitive searches (even in the \nustar era). Only two potential features have
been seen from the ULX population to date and these limited results paint something of
a mixed picture (\citealt{Brightman18, Walton18crsf, Middleton19}). Other potential
means of estimating the magnetic field strengths in ULXPs are thus  clearly
of interest.

One thing a number of the known ULXPs have in common is that they show
strong long-term variability, sometimes including transitions to "off-states" in which their X-ray flux drops by
orders of magnitude  (e.g., \citealt{Motch14nat,
Earnshaw16, Brightman16m82a}). One possible explanation for these events is that
they represent transitions to the propeller regime, in which the magnetic field of the
neutron star (temporarily) acts as a barrier to accretion, resulting in a
precipitous drop in the observed X-ray flux (\citealt{Illarionov75, Tsygankov16, Earnshaw18}). If this is correct,
then they may offer an independent means to estimate the dipolar magnetic fields of these
sources. However, the nature of these events is not yet clear, and may differ for
different systems and events. For example, in M82 X-2 X-ray monitoring suggests
that low-states are related to the $\sim$64\,d super-orbital period seen in that system
(\citealt{Brightman19m82}), which would be somewhat challenging to reconcile with a 
propeller-based interpretation. Furthermore, other authors have suggested that these
low-flux periods may be related to obscuration, instead of a cessation of accretion
(\citealt{Motch14nat, Vasilopoulos19, Fuerst21}).


Among the known ULXPs \ngc is the most luminous, with an astonishing peak
luminosity of $L_{\rm{X}} \sim 10^{41}$\,\ergs\ ($\sim$500 times the Eddington
luminosity for a standard neutron star assuming a distance of $d=17.1$\,Mpc; \citealt{Israel17, Tully16}), making it a case of particular interest.
Coordinated observations with \xmm and \nustar show that when the source is bright, it shows broadband spectra that are well described with a combination of a super-Eddington accretion disk that contributes thermal emission below $\sim$10\,keV and an accretion column that dominates at higher energies (\citealt{Walton15, Walton18ulxBB, Fuerst17ngc5907}).
Studies of the short-timescale evolution of its pulse
period revealed a potential orbital period of $\sim$5\,d (\citealt{Israel17}; however, this period is not yet confirmed), and when
in its ULX state the source is also known to exhibit a $\sim$78\,d X-ray period
(\citealt{Walton16period}), which is therefore likely to be super-orbital in nature. \ngc is
also one of the ULXPs that is known to exhibit intermittent off-states
(\citealt{Walton15}). However, unlike the case of M82 X-2, the low-flux periods in
\ngc are not related to its super-orbital period and so, they could plausibly be related
to the propeller transition.

As reported by \citet{Israel17}, the earliest measurement of pulsations from \ngc are from February 2003, with a spin period $P$ around 1.43\,s. Eleven years later, the pulsar had spun-up to $P\approx1.14$\,s \citep{Israel17}. Assuming a constant spin-up over this period implies a rate of change of the pulse period $\dot P= -8.1\times10^{-10}$\,s\,s$^{-1}$. This long-term spin-up indicates that the pulsar was not at spin equilibrium and that the high mass accretion rate also resulted in a large increase of angular momentum. Similar long-term spin-ups are observed for other ULXPs, like NGC\,7793 P13 \citep[$\dot P \approx -4\times10^{-11}$\,s\,s$^{-1}$,][]{Fuerst21} and NGC\,300 ULX1 \citep[$\dot P \approx -4\times 10^{-10}$\,s\,s$^{-1}$,][]{Vasilopoulos19}.

As the spin-up is driven by transfer of angular momentum through accretion, this can also be theoretically used to measure the magnetic field strength. This fact has been discussed in detail by the seminal papers by \cite{ghosh79a} and \cite{ghosh79b}, which were subsequently updated by \citet{Wang95}. However, those calculations are based on various simplifications, in particular, they assume a geometrically thin, optically thick accretion disk as expected for sub-Eddington accretion rates (\citealt{Shakura73}). In ULXPs, however, we expect the accretion disk to be geometrically thick due to their super-Eddington accretion rates \citep{Shakura73, Abram88}. Nonetheless, using the \citet{ghosh79b} model for ULXPs NGC\,7793~P13 and NGC\,300 ULX1, magnetic fields around $10^{12}$\,G were implied \citep{Fuerst16p13, Carpano18}, which is very well in line with the magnetic fields observed in high-mass X-ray binaries in our own galaxy \citep[see, e.g.,][]{staubert19a}. 

For \ngc, much higher magnetic fields have been postulated based on the extreme luminosity, with a multipolar magnetic field component as high as a few $10^{14}$\,G \citep{Israel17}. More recently, using an updated description of the coupling between the magnetic field and the accretion flow intended to be more suitable for a super-Eddington disk, \citet{gao21} estimated a maximal field between 2--3\,$\times 10^{13}$\,G. These authors, however, find that the observations presented by \citet{Israel17} are also consistent with a field around $6\times10^{11}$\,G.

We are therefore in need of another approach to constrain the magnetic field. One possibility is to study the expected spin-down during a low flux phase, in which accretion might be inhibited due to the fast rotation of the neutron star. In this regime, the spin-down torque onto the neutron star is dominated by the magnetic field interacting with the residual accretion disk \citep{parfrey16a}. \ngc entered a suitable low-flux state in mid-2017, which, broadly speaking, lasted until mid-2020, with brief episodes of higher flux in between. In this paper, we discuss the pulse period evolution before, during, and after this extended off-state, as observed with \xmm, and show that the pulsar spun-down significantly during the off-state.

The remainder of the paper is structured as follows. In Sect.~\ref{sec_red}, we describe the data used and extraction methods. In Sect.~\ref{sec_timing} we discuss the flux and period evolution and describe our pulsation search in detail. In Sect.~\ref{sec_res}, we  discuss our results and use the measurements to estimate the magnetic field. Section~\ref{sec_conc} provides a conclusion, along with a summary of our main results and an outlook for future measurements.

\begin{figure*}
\begin{center}
\includegraphics[width=0.95\textwidth]{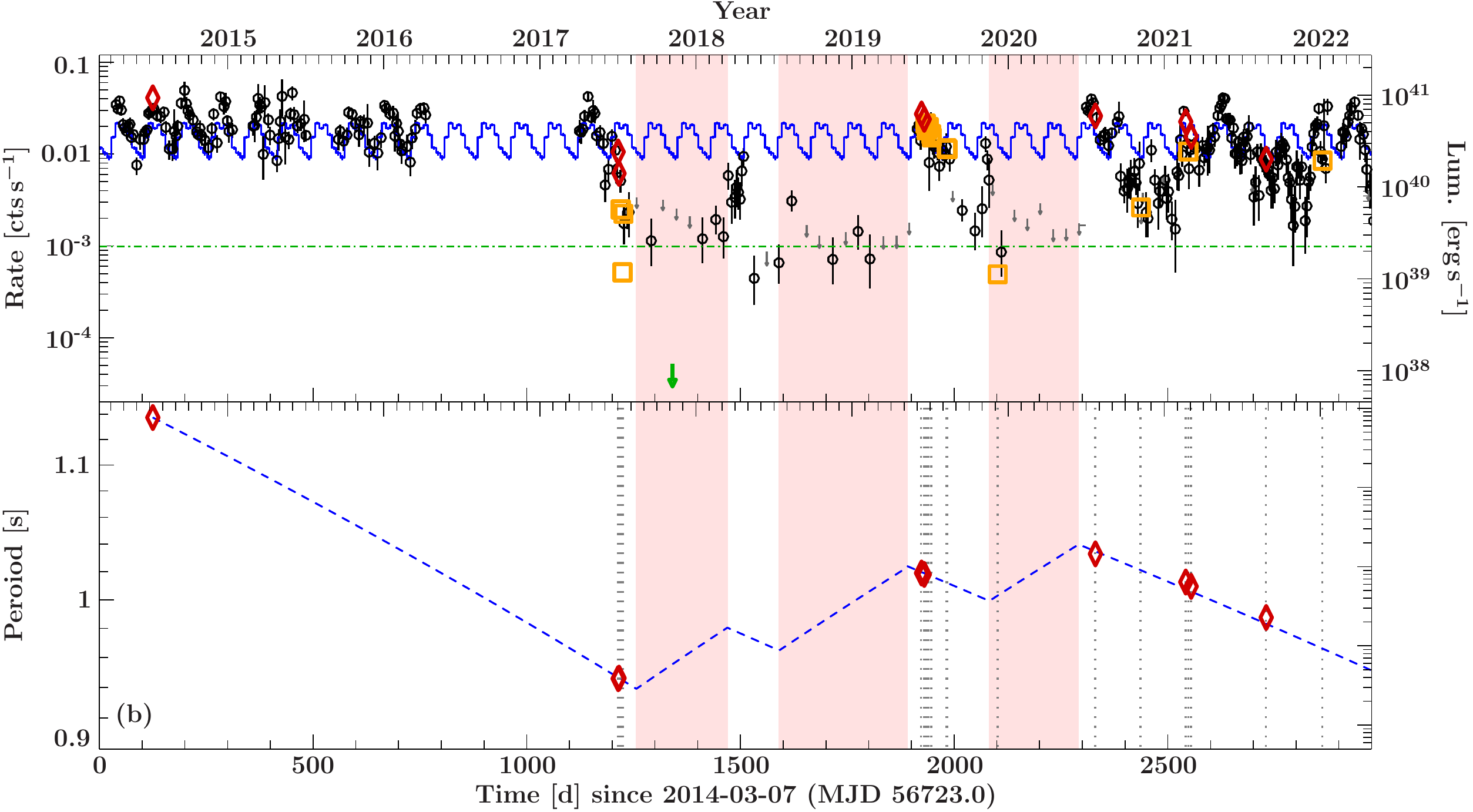}

\end{center}
\vspace*{-0.3cm}
\caption{
Flux and period evolution of \ngc between 2014 to 2022. 
\textit{Top:}  \swift\ XRT light curve (0.3--10.0\,keV). 
 In green the upper limit for the point source luminosity as measured with \chandra \citep{belfiore20a} is shown, using the right $y$-axis. The \xmm luminosities are shown as red diamonds and orange squares for observations with and without detect pulsations, respectively, also using the right-hand $y$-axis.
The horizontal green line indicates the estimated propeller luminosity based on a magnetic field strength of $B=2.6\times10^{13}$\,G (see Table~\ref{tab:bsummary}). 
The blue curve shows an extrapolation of the
78\,d X-ray period seen from \ngc\ during its ULX state (\citealt{Walton16period}).
The shaded pink areas indicate where the source was conceivably in the low state and spinning down.
\textit{Bottom:} Pulse period measurements, as listed in Table~\ref{tab_obs} and \citet{Israel17}. The gray vertical dotted lines indicate times of observations with \xmm. The blue dashed line shows a possible fiducial model of epochs of spin-up during bright states interspersed with spin-down during off-states. 
This line is only a suggestion for the evolution of the period.
For details see text. 
}
\label{fig_longlc}
\end{figure*}

\section{Observations and data reduction}
\label{sec_red}

\subsection{\swift}

\ngc has been extensively monitored by the \swiftng\ and, in particular, the X-ray telescope (XRT; \citealt{SWIFT_XRT}). The
\swift\ XRT light curve (Figure \ref{fig_longlc}) is extracted using the
standard online pipeline (\citealt{Evans09}, primarily using time bins
of four days while the source was bright (following \citealt{Walton16period}).
The exception to this is when \ngc drops its flux significantly. During
these periods, we revert to an approximately monthly binning. The average count
rates for these broader bins are determined by extracting XRT images and
exposure maps integrated across them, also using the online XRT pipeline,
and calculating the net count rate observed using a circular source region
of radius $10''$ and a much larger neighboring region to estimate the
background. Uncertainties are calculated following \cite{Gehrels86}.
Where the source is not detected, an upper limit at the 3$\sigma$-level is calculated
using the method outlined by \cite{Kraft91}. These rates and upper limits
are then corrected for the fraction of the XRT PSF that falls outside of
the source extraction region. The \swift monitoring snapshots do not provide enough photons to search for or measure the pulse period.

\subsection{\xmm}

In order to understand the nature of the strong variability seen by \swift, we executed a series of \xmm\ observations over the last few years designed to monitor the spin period of the source, both across and after the extended low-flux period seen from $\sim$2017--2020. In addition, \xmm also performed a series of earlier observations extending as far back as 2003. Details of these observations are given in Table~\ref{tab_obs}.

For all of these \xmm observations (see Table \ref{tab_obs}), the
EPIC detectors (\epicpn, \epicmos; \citealt{XMM_PN, XMM_MOS}) were operating in
full-frame mode. We therefore focus on the data taken by \epicpn\ here (temporal
resolution of 73.4\,ms), as the \epicmos\ detectors do not have sufficient timing
capabilities to probe the $\sim$1\,s spin period of the neutron star (temporal
resolution of $\sim$2.6\,s). Data reduction was carried out with the \xmm\ Science
Analysis System (\sas v19.1.0), largely following standard procedures. The raw observation data files were processed using \epchain and the cleaned event files corrected to
the solar barycenter using the DE200 solar ephemeris with \barycen. Source
light curves were then extracted on the time resolution of the \epicpn\ detector with
\xmmselect. We typically used circular source regions
of radius 25--30$''$, depending on the brightness of
\ngc, although for some of the extremely faint observations, even smaller regions were occasionally used (ranging down to $\sim$15$''$). As recommended we only considered single and double
patterned events.

In all cases, the background was estimated from larger regions of blank sky on the same detector as the source region. 
Given the large number of \xmm\ observations considered, it is unsurprising to notice that a broad range of background activity is seen among them; some suffer from severe flaring, some show brief, modest periods of enhanced background and some show a stable background level throughout. In order to determine whether additional background filtering is required -- and, if so, then subsequently establishing the level of background emission that is acceptable -- we utilized the method outlined in \cite{Picon04}.
This determines the background level at which the signal-to-noise (S/N) of the source is maximized. We make this assessment for each observation individually, and maximize the S/N over the  0.3--10.0\,keV band. For cases where additional background filtering is required, custom good-time-intervals are generated to exclude background levels above that which gives the maximum S/N for the source.

\begin{table*}
  \caption{Details of the \xmm\ observations of \ngc\ considered in this work. The uncertainties given for the pulse period and its derivative are statistical only and are dominated by the larger uncertainties of the orbital ephemeris ($\approx0.2$\,ms). The luminosity is based on the absorption corrected flux in the 0.3--10\,keV band and the pulsed fraction is given for the 1--10\,keV energy band.}
\vspace{-0.5cm}
\begin{center}
\begin{tabular}{l l l l l l l l}
\hline
\hline
\\[-0.25cm]
OBSID &  Start & Exposure & Luminosity  & $P_{\rm{spin}}$  & $\dot P_{\rm{spin}}$ & PF  \\
&   Date & (ks) & ($10^{39}$\,erg\,s$^{-1}$) &  (ms) & ($10^{-10}$\,s\,s$^{-1}$) &  \% \\
\\[-0.3cm]
\hline
\hline
\\[-0.2cm]

0145190201\tablefootmark{a}  &   2003-02-20 &  41.3 &   $81\pm4$ &  $1427.76^{+0.11}_{-0.09}$ &  $-90^{+40}_{-50}$ &  $19.4\pm2.5$ \\

0729561301\tablefootmark{a}  &   2014-07-09 &  42.0 &   $93.4^{+2.5}_{-2.8}$ &  $1137.43^{+0.06}_{-0.07}$ &  $-50\pm28$ &  $21.5\pm2.3$ \\

0804090301  &   2017-07-02 &  33.5 &   $24.1^{+2.4}_{-2.6}$ &  $945.79^{+0.08}_{-0.05}$ &  $23^{+24}_{-29}$ &  $19\pm5$ \\

0804090401  &   2017-07-05 &  36.0 &   $14.1\pm1.3$ &  $946.17^{+0.04}_{-0.08}$ &  $0^{+37}_{-20}$ &  $30\pm6$ \\

0804090501  &   2017-07-08 &  40.0 &   $5.7\pm0.9$ &  -- &  -- &  $<36.0$ \\

0804090701  &   2017-07-12 &  40.0 &   $1.2^{+0.5}_{-0.8}$ &  -- &  -- &  $<44.0$ \\

0804090601  &   2017-07-15 &  37.5 &   $5.1^{+0.6}_{-1.0}$ &  -- &  -- &  $<43.0$ \\

0824320201  &   2019-06-12 &  59.7 &   $62.4^{+1.9}_{-2.1}$ &  $1019.142^{+0.026}_{-0.046}$ &  $-29^{+16}_{-9}$ &  $20.1\pm2.3$ \\

0824320301  &   2019-06-19 &  49.1 &   $54.7^{+2.0}_{-2.1}$ &  $1018.10^{+0.05}_{-0.07}$ &  $-13^{+23}_{-18}$ &  $13.5\pm2.8$ \\

0804090801  &   2019-06-22 &  38.7 &   $48.8^{+2.7}_{-2.6}$ &  -- &  -- &  $<13.0$ \\

0804090901  &   2019-06-24 &  40.2 &   $41.7^{+1.7}_{-1.9}$ &  -- &  -- &  $<16.0$ \\

0824320401  &   2019-06-26 &  56.4 &   $42.9^{+1.6}_{-1.7}$ &  -- &  -- &  $<12.0$ \\

0804091001  &   2019-06-29 &  46.0 &   $44.9^{+1.7}_{-1.8}$ &  -- &  -- &  $<15.0$ \\

0804091101  &   2019-07-05 &  34.8 &   $37.5^{+1.6}_{-2.1}$ &  -- &  -- &  $<19.0$ \\

0804091201  &   2019-07-06 &  37.3 &   $34.7\pm1.9$ &  -- &  -- &  $<18.0$ \\

0851180701  &   2019-08-10 &  56.2 &   $26.5^{+0.8}_{-1.3}$ &  -- &  -- &  $<18.0$ \\

0851180801  &   2019-08-12 &  60.8 &   $25.7^{+0.8}_{-0.9}$ &  -- &  -- &  $<17.0$ \\

0824320501  &   2019-12-08 &  43.5 &   $1.12^{+0.21}_{-0.39}$ &  -- &  -- &  $<31.0$ \\

0824320601  &   2020-07-23 &  44.5 &   $59.5^{+2.1}_{-2.2}$ &  $1032.94^{+0.08}_{-0.05}$ &  $-23^{+21}_{-33}$ &  $13.0\pm2.7$ \\

0824320701  &   2020-11-06 &  34.5 &   $6.0^{+0.4}_{-1.0}$ &  -- &  -- &  $<19.0$ \\

0884220201  &   2021-02-20 &  54.7 &   $51.3^{+2.5}_{-2.6}$ &  $1012.620^{+0.025}_{-0.033}$ &  $-36^{+10}_{-9}$ &  $16.9\pm2.2$ \\

0884220301  &   2021-02-26 &  61.4 &   $24.4^{+1.1}_{-1.2}$ &  -- &  -- &  $<16.0$ \\

0884220401  &   2021-03-04 &  59.0 &   $35.5^{+2.0}_{-2.2}$ &  $1009.563^{+0.027}_{-0.035}$ &  $-31^{+11}_{-9}$ &  $12.9\pm2.3$ \\

0891801501  &   2021-08-27 &  59.0 &   $20.1^{+1.1}_{-1.2}$ &  $987.558^{+0.030}_{-0.036}$ &  $-11^{+10}_{-11}$ &  $30\pm4$ \\

0893810301  &   2022-01-06 &  40.0 &   $19.3^{+1.1}_{-1.5}$ &  -- &  -- &  $<17.0$ \\
\\[-0.15cm]
\hline
\hline
\end{tabular}
\label{tab_obs}
\end{center}
\vspace{-0.3cm}
 \tablefoot{\tablefoottext{a}{Pulse period based on values presented by \citet{Israel17}.}}
\end{table*}

\section{Timing analysis}
\label{sec_timing}

\subsection{Light curve and flux evolution}
\label{susec_lc}

The evolution of \ngc from early 2014 to early 2022 as seen by \swift is shown in Figure~\ref{fig_longlc}. Before 2014, no dense monitoring of the source is available, although there are a small number of observations with \xmm, \swift,\ and \nustar\ prior to this.
The source faded significantly in 2017 -- to the extent that it became
challenging to detect with \swift and only returned to full activity briefly in $\sim$May 2019; we note that there was also a brief re-brightening in $\sim$March 2018, but the source did not fully return to its "normal" ULX state. This low-flux period ended in mid 2020, and since then \ngc has been detected in almost every \swift snapshot obtained. Despite the broad recovery, comparing the behavior before 2017 and after 2020, we can see that the source is still exhibiting a much larger variability in the more recent data, with flux changes of up to a factor of $\sim$10 within a few months. 

Previously, \ngc was also in a low flux state between 2012--2014 (see \citealt{Walton15}), but its duration and exact luminosities are unclear due to the lack of monitoring. Between 2003--2012 the source seems to have largely been in a bright on-state, as seen by \swift, \xmm, and \chandra \citep{Israel17}. 

Fortuitously, during the initial decline in flux in 2017 a series of five \xmm observations were taken, allowing us to obtain a precise measurement of the neutron star spin just before the low state (Table~\ref{tab_obs}).
The activity period in early 2019 was
sufficiently bright and long-lived for us to trigger another series of ten \xmm
observations in order to compare the properties of \ngc in 2017 and 2019 and
determine how the source has evolved across the extended low-flux period in between.
Furthermore, another series of seven \xmm\ observations of \ngc\ have been performed since its more persistent recovery in mid-2020, allowing for further comparisons of the recent behavior with that seen in 2019 as well as prior to 2017.
In particular, we focus on timing the pulsar and tracking the evolution of its spin period in this work.


\subsection{Pulsations}
\label{susec_pulsations}
As a simple first step in the timing analysis we performed a Fast-Fourier Transform (FFT) on all full-band EPIC-pn light curves, extracted with a time resolution of 73.4\,ms. This approach revealed significant pulsations at $\sim$0.9812\,Hz in ObsID 0824320201 (2019-06-12). In all other observations, no significant signal was found using this basic analysis.

Based on the variation of the pulsed fraction as function of energy available in the literature (\citealt{Israel17}), we find that below 1\,keV the pulsations are barely detectable. We therefore subsequently filtered all data on energies $>1$\,keV for the following analysis to increase the S/N. Following this additional filtering, as a next step, we performed a more in-depth pulsation search using an accelerated Fourier method, which searches a grid in the frequency, $\nu,$ and frequency-derivative $\dot \nu$ space. We used the implementation \texttt{HENaccelsearch} from the HENDRICS software package \citep{hendricsref} and performed the search between $\nu$=0.9--1.2\,Hz. This analysis revealed good pulsation candidates in ObsIDs 0804090301, 0804090401, 0824320201, 0823430401, and 0891801501, with a detection significance of $>3\sigma$.

For each of these observations, the best $\nu$-$\dot \nu$ combinations from this search were then analyzed in more detail, with a grid search using epoch folding \citep{Leahy83}. The data were searched in temporal space across a range of $\Delta P =  \pm0.3$\,ms and $\Delta \dot P \pm  = 1\times10^{-8}$\,s\,s$^{-1}$, respectively, centered on the best values from \texttt{HENaccelsearch}.
We oversampled the search in $P$ by a factor of 5 compared to the number of independent frequencies and used the same number of grid points also in $\dot P$ space. 
This analysis is performed with ISIS \citep{ISIS}, following a similar procedure to that described in \citet{Fuerst16p13}.
Uncertainties on $P$ and $\dot P$ are given as the value where $\chi^2$ has dropped to half its maximum value, namely, an FWHM width based on the $\chi^2$ landscape in the $P$--$\dot P$ plane.

The results of this search are given in Table~\ref{tab_obs} and shown in the bottom panel of Fig.~\ref{fig_longlc}. 
It is clear from these results that the neutron star was spinning significantly slower in 2019 compared to 2017 and again slightly slower in 2020 compared to 2019. After mid 2020, \ngc\ can be seen to have restarted its steady spin-up trend.

We then also performed the same epoch folding search for all ObsIDs where \texttt{HENaccelsearch} did not find a significant signal. The searches were centered on a period obtained from linear interpolation or extrapolation between the neighboring pulsation detections. We increased the search range to $\Delta P = 3$\,ms and $\Delta \dot{P} = 5\times10^{-8}$\,s\,s$^{-1}$ and again oversampled the $P$ space by a factor of 5. Larger grids are computationally prohibitive for this kind of brute force method. We did not find significant pulsations in any of these observations.

The pulse periods we give here are not corrected for the orbital motion of the neutron star. We chose not to perform this correction given the large uncertainties on the current ephemeris \citep{Israel17}. However, taking the values given by \citet{Israel17} and their respective uncertainties, we find that the maximal influence of their orbital solution on the measured spin-period is on the order of 0.2\,ms. This is smaller than the difference in spin period that we find between the two 2019 observations -- and much smaller than the differences between the 2017 and 2019 observations.

\subsection{Pulsed Fraction}

Where pulsations have been detected, we calculate the pulsed fraction based on the pulse profile (PP) as
\begin{equation}
PF = \frac{\max (PP) - \min (PP)}{\max (PP) + \min(PP).}
\end{equation}

\noindent{}For observations without detected pulsations, we instead calculate the upper limits on the pulsed fraction. To do so, we simulate a series of event lists with the same GTIs and average count rate as the real data but adding sinusoidal pulsations with an increasing pulsed fraction. The period assumed for these pulsations is based on the closest measured periods and is determined by linearly interpolating the evolution of the pulse period between these measurements to the time of the observation in question. For each pulsed fraction, we simulated 50 individual event lists and performed the same search for pulsations as applied to the real data. We consider pulsations to be detected in our simulated datasets when we find a peak at the 99.9\% significance level or above. Starting from 5\%, the pulsed fraction is increased in steps of 0.1\% up to the point where the fraction of simulations that return significant pulsations increases above 90\% (i.e., the pulsations are recovered in at least 45 simulated event lists). We take this pulse fraction to be the upper limit for the observation in question (Table~\ref{tab_obs}).

\section {Results \& discussion}
\label{sec_res}

\subsection{Magnetic field based on spin-up strength}
\label{susec_spinup}

As seen in Fig.~\ref{fig_longlc}, we can identify two epochs of relatively stable spin-up, the first one between 2014 and 2017 (starting around MJD~56847) and the second one after mid 2020 (starting around MJD~59053). In addition, based on the results presented in \citet{Israel17}, the source was also spinning up on average between 2003 to 2014 (starting around MJD~52690). However, we do not know how stable the spin-up and X-ray flux was between 2003 and 2014, due to the lack of regular monitoring.
Using these three spin-up episodes, we can try to estimate the magnetic field strength based on theoretical calculations describing how the accreted matter couples to the neutron star via the magnetic field, as well as how much angular momentum is transferred by this coupling.
Throughout the remainder of this paper, we assume a neutron star radius $R = 10^6$\,cm and a neutron star mass of $M=1.4\,\msun$, unless otherwise noted.

To understand the accretion torque on the neutron star, it is important to analyze the relationship between a number of important radii at which significant changes in the geometry or the dominating physical processes occur. These are detailed below. 
    
    The corotation radius, $R_c$, is the radius at which the Keplerian orbital speed is equal to the rotational speed of the pulsar. High accretion rates can only occur if the accreted material couples inside this radius to the magnetic field. The corotation radius is given by:
    \begin{equation}
    \begin{split}
        R_c & = \left(\frac{G}{4\pi^2}\right)^{1/3} \, P^{2/3} \, M^{1/3} ~ \mathrm{cm} \\
        & \approx 1.7\times10^{8} ~ P^{2/3} \, M_{1.4}^{1/3} ~ \mathrm{cm}
    \end{split}
    \label{eq:rc}
    .\end{equation}
    
    Next, there is the magnetospheric radius, $R_M$, which is the radius at which the magnetic field dominates over the ram pressure of the accreted material and within which the material has to follow the magnetic field lines. This is equivalent to the inner radius of the accretion disk, and is related to the Alfv\'en radius computed for spherical accretion with a factor $\xi$:
    \begin{equation}
    R_M = \xi \mu^{4/7} \left(2GM \mdotin^2\right)^{-1/7} ~ \mathrm{cm.}
    \label{eq:magnetosphere}
    \end{equation}
    
    Then we have the spherization radius, \rsph, the radius at which the accretion becomes locally super-Eddington and geometrically thick. Following \citet{Shakura73}, this is given by:
    \begin{equation}
    \rsph = 3GM\mdottot \left(2 \ledd\right)^{-1} ~ \mathrm{cm.}
    \end{equation}

\noindent{}In the above expressions, $P$ is the pulse period in seconds, $M_{1.4}$ is the neutron star mass in units of 1.4\,\msun, $\mu$ is the magnetic moment, $G$ is the gravitational constant, \mdotin is the accretion rate at the inner edge of the accretion disk, $\mdottot$ is the total mass transfer rate, and \ledd is the Eddington luminosity. For a dipolar magnetic field, $\mu = B R^{3}$, where $B$ is the magnetic flux density of the dipolar field. 

The problem in calculating these radii lies in the fact that they depend on $\mdottot$ as well as \mdotin, which, in principle, can be inferred from the observed luminosity, $L$. However, the relation between \mdotin and the luminosity itself depends on the relative location of \rsph, $R_M$, and $R_C$.

Within \rsph, the outflows will ensure that the local Eddington luminosity is not exceeded, leading, at the same time, to the formation of a funnel that will go on to collimate the observed emission. As discussed by \citet{King09}, for high accretion rates in the super-Eddington regime, we have: 
\begin{equation}
    L = \frac{\ledd \left(1 + \ln \dot m\right)}{b} 
    \label{eq:LumSuperEdd}
    ,\end{equation}

\noindent{}where $\dot{m} = \mdottot / \mdotedd$, \mdotedd is accretion rate that corresponds to the Eddington limit and $b$ is the beaming factor, which can be approximated as $b \approx 73/\dot m^2$.
High accretion rates in this case mean $\dot m > 8.5$ \citep{King17ulx}, which is the case for all observations of \ngc considered here.

The estimated mass accretion rate $\mdottot$ from Eq.~\ref{eq:LumSuperEdd} can be used to calculate \mdotin, which is the relevant accretion rate for determining $R_M$. For the same reasons as discussed for the luminosity, \mdotin is rising slower than $\mdottot$ inside of \rsph to guarantee that the disk does not locally exceed its Eddington limit.
In particular, following \citet[][\textit{cf} \citealt{Erkut19}]{Shakura73}:
\begin{equation}
\dot{M}_\text{in} =
\begin{cases}
\mdottot (R_M/\rsph), \quad & \text{for } R_M<\rsph, \\
\mdottot, \quad & \text{for } R_M>\rsph.
\end{cases}
\label{mdot}
\end{equation}

Finally, in order to calculate $R_M$, we have to determine $\xi$, the conversion factor for the Alfv\'en radius calculated under the assumption of spherical symmetry. The value of $\xi$ depends strongly on our assumptions of how the accreted material couples and interacts with the magnetic field, depending on the magnetic field geometry. \citet{gao21} distinguish two cases of the ratio between the azimuthal and vertical B-field strength which describe different physical regimes of the coupling between the accretion disk and the magnetosphere \citep[see also][]{Wang95}. These two cases lead to different expressions of the total dimensionless torque $n(\omega),$ where $\omega$ is the so-called  fastness parameter. The exact expressions of these equations can be found in \citet{gao21}. Based on the description of the transfer of angular momentum onto a rigidly rotating neutron star, we can solve for $\omega$. These equations are at least quadratic in $\omega$, leading to two solutions for each case. We can then solve for the magnetic field, using the following equations \citep{gao21}.\\
\noindent For $R_M \geq \rsph$:
\begin{equation}
    B = 1.371\times10^{12} \xi^{-7/4} \omega^{7/6} M_{1.4}^{5/6} R_6^{-3} \mdot^{1/2} P^{7/6}\,\text{G.} 
\end{equation}
And for $R_M\leq \rsph$:
\begin{equation}
    B = 1.234\times10^{13} \xi^{-7/4} \omega^{3/2} M_{1.4}^{1/2} R_6^{-3} P^{3/2}\,\text{G.} 
\end{equation}
Here, $\xi$ can be approximated with:
\begin{equation}
    \xi \approx 2^{-1/7} \left( 1-\omega\right)^{2/7}
    \label{eq:xigao}
,\end{equation}
which is the same regardless of whether the magnetospheric radius is larger or smaller than the spherization radius. Also,  $R_6$ is the neutron star radius in units of $10^6$\,cm.

Given the high luminosity of \ngc, we expect that  the spherization radius tends to be outside of $R_M$. However, for certain combinations of magnetic field strengths and coupling assumptions, we can also find that this may not be the case. For these potential solutions, the magnetic field is so strong that it disrupts the accretion disk before it can become locally super-Eddington.  
We have up to four $B$-field estimates in each case, however, not all estimates fulfill the assumptions (i.e., for an overly strong magnetic field, $R_M$ might be outside of \rsph, making the solutions for $R_M \leq \rsph$ invalid in this case ).
In Table~\ref{tab:bfields}, we list all possible solutions for the three spin-up epochs, that is, for each set of input values, $L$, $P$, and $\dot P$.

As can be seen in Table~\ref{tab:bfields}, we find roughly similar magnetic field strengths for epochs 2 and 3, but a different value for epoch 1. The long baseline for estimating $\dot P$ in this epoch and the lack of flux monitoring can explain this difference: the average $\dot P$ and $L$ are likely to be inaccurate, leading to unreliable estimates.
Epochs 2 and 3 show a particularly good agreement for the high $B$-field solution, with a variation of only $<10\%$ for both cases, while the difference for the low $B$-field solution is around $\sim50\%$. 
Nonetheless, without further information, we cannot determine which solution describes the true nature of the system better.

If we assume that the neutron star has a $B$-field of $2.0\times10^{13}$\,G (roughly the average of the high-B solutions based on epochs 2 and 3), we can also estimate the average luminosity during epoch 1, which comes out to about $7\times10^{39}$\,\ergcms. While this value may be a bit on the low side, it could be realistic, given that \ngc did seemingly exhibit low fluxes  throughout much of 2013 (\citealt{Walton15}). In this case, $R_M > \rsph$, as the luminosity is very low compared to the magnetic field.

Even with this lower estimated luminosity for epoch 1, the low $B$-field solution does not agree with the other epochs. This discrepancy is mainly due to the fact that the $B$-field estimate is independent of the luminosity as long as $R_M < \rsph$, as the mass accretion rate at the inner accretion disk edge is regulated by the Eddington limit.

\subsection{The off-state}
\label{susec:offstate}

In the simplest picture, an accreting neutron star will enter the propeller (or centrifugal inhibition) regime as soon as the corotation radius is inside the magnetospheric radius ($R_M > R_c$). At the magnetospheric radius, matter couples to the magnetic field and is forced to rotate at the angular speed of the neutron star. In the case of $R_M > R_c$, this rotation is faster than the Keplerian speed and therefore the matter experiences a net outward force, all but halting direct accretion. The accretion luminosity will then drop by orders of magnitude, however, residual accretion at the magnetosphere might still be present, that is, the source does not have to appear completely off \citep[see, e.g.,][]{corbet96a}.

We can find the critical mass accretion rate for the transition to propeller regime, $\mdotincrit{}$ by equating $R_c = R_M$ (using Eqs.~\ref{eq:rc} and \ref{eq:magnetosphere}), leading to:

\begin{align}
      \mdotincrit &= \left(\frac{1}{4\pi^2}\right)^{-7/6} \, \frac{1}{\sqrt{2}}  G^{-5/3}  P^{-7/3} M^{-5/3}  \xi^{7/2} \mu^{2},\\
     &\approx 4.69\times10^{13} \cdot P^{-7/3} M^{-5/3} \xi^{7/2} \mu^{2} \,\text{g\,s}^{-1}.
\end{align}

Based on this calculation, we give the critical luminosity for each epoch and case in Table~\ref{tab:bfields}, which is $L_\text{crit} = \mdotincrit \epsilon c^{2}$, where $\epsilon = 0.15$, the accretion efficiency of a typical neutron star on its surface.  Given that the critical luminosity depends strongly on the magnetic field, the estimates differ more strongly, for instance, by a factor of 2 for the low $B$-field case (again, only comparing epochs 2 and 3). 

Observationally, we can also estimate the luminosity of a possible onset of the propeller effect from the \swift/XRT light curve, by finding the flux at which the source drops below the detection limit of XRT. To estimate this luminosity we transferred the \swift count-rate to flux and luminosity based on the contemporaneous \xmm data. We use a conversion factor of:
\begin{equation} 
C =  \frac{ \mathcal{F(\xmm)}}{\text{Rate(\swift)}}  = 6.516\times10^{-11}\, \frac{\ergcms{}}{\text{cts\,s}^{-1}} \quad. \end{equation}

As can be seen on the right-hand $y$-axis of Fig.~\ref{fig_longlc}, the source drops below the detection limit of \swift around 1--2$\times10^{39}\,\ergs$. \xmm detects the source around $1\times10^{39}$\,\ergs during the lowest flux states, but no pulsations were found. This luminosity is therefore a good upper limit for the onset of the propeller effect and we refer to it as the propeller luminosity. That value is consistent with the fact that pulsations were only detected down to $14\times10^{39}\,\ergs$ in July 2017 (ObsID 0804090401).

While XRT is not very sensitive, we know that accretion was heavily suppressed in November 2017, due to a deep \chandra observation. This \chandra observation revealed a diffuse nebula around the ULXP and provided a stringent upper limit of $L<1.2\times10^{38}\,\ergs$ for the point source \citep{belfiore20a}. We include this upper limit in Fig.~\ref{fig_longlc}.

The observational limit of $1\times10^{39}\,\ergs$ for the propeller luminosity is a little bit below the luminosity implied by the highest estimated $B$-field from the spin-up (Tab.~\ref{tab:bfields}), namely, $L_\text{crit} = 2.2\times10^{39}\,\ergs$ for a magnetic field of $B = 2.6\times10^{13}$\,G (orange dotted-dashed line in Fig.~\ref{fig_longlc}).

It is worth stressing that the initial decline in flux seen during the dense \xmm monitoring in July 2017 is not itself related to the propeller transition in our interpretation. Rather, this must be related to some other physical effect that resulted in a gradual reduction in the overall accretion rate through the disk and the disk itself is also expected to make a significant contribution in the XRT band \citep[see, e.g.,][]{Walton18}. For example, density waves in the accretion disk might have reduced the reduced the inner accretion rate or perhaps the mass transfer rate from the stellar companion decreased slightly. We consider this initial stage of the decline to be similar to the behavior seen in 2021 and 2022, where the source also shows significant flux variability without fully entering an off-state. However, in 2017 the flux continued to drop even further, to the point where it reached our proposed propeller luminosity, resulting in a transition to the propeller regime. This transition would naturally explain the very stringent upper limit observed in November 2017 by \chandra. However, as this observation occurred about 4 months later, we cannot make any firm statement with regards to how rapidly the actual propeller transition would have occurred with these data and, unfortunately, the XRT does not have the sensitivity to meaningfully shed light on this issue either (the source is already not detected by the XRT at our propeller luminosity). We note, however, given the rise by two orders of magnitude (or more) in flux in only four days, as seen by \citet{Walton15} using \xmm and \nustar, these observations (taken in 2013) may provide the most stringent constraints on this issue.

We can also follow a more detailed description of the super-Eddington accretion disks put forward by \citet{Chashkina17, Chashkina19}. Their model in particular takes advection of material in the accretion disk into account, as well as illumination of the disk by the luminosity emitted close to the compact object. 

For a radiation-dominated disk, they find an updated description of the inner disk radius, which does not depend on the mass accretion rate \citep[Eq. 61 in][]{Chashkina17}:
\begin{equation}
    R_M = \left( \frac{73}{24} \alpha \lambda \mu_{30}^2\right)^{2/9} \cdot \frac{GM}{c^2}\,\text{cm}\quad{,}
\end{equation}
where $\alpha$ is the viscosity of the disk, based on the standard \citet{Shakura73} description and $\lambda \approx 4\times10^{10} M_{1.4}^{-5}$. We assume $\alpha = 0.1$. Here $\mu_{30}$ is the magnetic moment of the neutron star in units of $10^{30}$\,G\,cm$^{-3}$ and $c$ is the speed of light.

Assuming that we reach the equilibrium spin-period just before entering the off-state, that is, a period of $P=0.946$\,s as measured in July 2017, we can equate $R_M = R_c$ and then estimate a magnetic field of around $6\times10^{13}$\,G. 
A similar estimate is found when following the description of  \citet{King17ulx}, who use the standard formula for the magnetospheric radius (Eq.~\ref{eq:magnetosphere}).

Based on our assumption that \ngc is spinning close to equilibrium and assuming that it entered the propeller state at the lowest measured luminosity where pulsations were still detected (in July 2017), we can rewrite the limit on the mass-accretion rate given Eq. 33 of \citet{Chashkina19} to estimate the B-field:

\begin{equation}
\mu^2_{30} = \frac{1}{1.8} \,\dot m\, \xi^{-7/2} M_{1.4}^{5/3} P^{7/3}\,\text{G}^2\,\text{cm}^{-6}\quad{.}
\end{equation}

Again, we use the measurements of July 2017 (ObsID 0804090401), that is, $P=0.946$\,s and based on Eq.~\ref{eq:LumSuperEdd}, $\dot m = 31.9$ and we approximate $\xi = 0.75$ (based on Eq.~\ref{eq:xigao}), we find a B-field of around $B\approx1.7\times10^{13}$\,G. This estimate is slightly lower due to the effects of irradiation and advection in this model. 

If we assume that \ngc{} entered a propeller state when it became undetectable by \swift, we would also expect that the source spins-down during this period. This behavior is indeed what we observe, as the spin period in 2019 is significantly slower than in 2017, just before the off-state.

However, the source rebrightened briefly in 2018, which likely means that it started accreting and spinning-up again. We do not have a pulsation measurement for this period, thus estimating the spin-down strength is not straight forward. In the bottom panel of Fig.~\ref{fig_longlc}, we indicate a possible spin-history of the source. We define regions (shaded pink) in which the source is off and spinning-down, thereby splitting up the data into seven time slices that alternate between spin-up and spin-down. 

In this model, we assume a spin-up strength of $\dot P = - 2.03\times10^{-9}$\,s\,s$^{-1}$ between 2014--2017 and a spin-up strength of $\dot P = -1.50\times10^{-9}$\,s\,s$^{-1}$ during all other spin-up episodes. For the spin-down, we assume a value of $\dot P = 2.25\times10^{-9}$\,s\,s$^{-1}$, which is  based on a spin-down estimate during the off-state in 2020.

This model is of course highly simplified and averages over long periods of time. It is possible that during the periods where the source was detected in XRT at $\sim10^{39}$\,\ergs, active accretion occurred and the source was spinning-up slightly (or at least, was not spinning down further) if $R_M \approx R_C$. However, the current data do not provide the required sensitivity and time resolution to model the spin history on shorter time-scales.

Theoretical estimates of the spin-down strength during the propeller state are difficult and depend on various assumptions of the interaction between the magnetic field and the residual surrounding matter \citep[e.g.,][]{davies79a, urpin98a, ikhsanov01a, dangelo10a}. Here we follow calculations presented by \citet{parfrey16a}, which were originally motivated by millisecond pulsars. In particular, these calculations assume that the spin-down torque is completely dominated by the torque exerted by the open field-lines and that there is no interaction between the field and the disk inside the co-rotation radius. Based on equation 18 of \citet{parfrey16a} we can write an estimate for the magnetic field: 
\begin{equation}
    B_\text{down} = \frac {2} {R^3} \frac {R_M} {\zeta} \sqrt \frac {I \dot P c}{P}\,\text{G}
,\end{equation}

where $I$ is the moment of inertia of the neutron star ($I = \frac25 M R^2$). The parameter $\zeta$ describes how efficiently the magnetic field lines are opened by the star-disk interaction. Here, we assume maximum efficiency, namely, $\zeta=1$, which implies that all field lines intersecting the disk are opened.

We measured an average spin-down $\dot P = 1.19\times10^{-9}$\,s\,s$^{-1}$ between mid 2017 to mid 2019, that is, over the two first off-states. The spin-down is likely a bit faster, given the short re-brightening of \ngc in 2018 (see Fig.~\ref{fig_longlc}). Nevertheless, with this value for $\dot P$, we find $B_\text{down} = 6.83\times10^{13}$\,G. This magnetic field strength is slightly higher than the largest value based on the spin-up calculations, but only by a factor of about 2--3. Given the significant number of simplifications and approximations going into this estimate, it is probably not surprising that we don't find a perfect agreement.
An overview of all our $B$-field estimates is given in Table~\ref{tab:bsummary}.

It is interesting to note that the source already showed a significant spin-down between two \xmm observations in 2017 (OBSIDs 0804090301 and 0804090401). These were taken while the source was on its decline into the off-state -- and were separated by only $\sim$3\,d. The evolution between these two observations implies $\dot P \approx 1.4\times10^{-9}$\,s\,s$^{-1}$, which is similar to what we measured between 2017--2019. We note, however that both observations are still significantly above the propeller luminosity that would correspond to a $10^{13}$\,G magnetic field, so there would appear to be an inconsistency here. Assuming an orbital period of 4.4\,d, the lower limit of the period proposed by \citet{Israel17}, we find that the observed spin-down is barely consistent with being due to the orbital motion within the uncertainties. However, the orbital period is not confirmed, so we cannot draw firm conclusions on the nature of this spin-down. 

Another possibility for the observed spin-down is that the source is entering the subsonic propeller (or magnetic inhibition) regime. In this regime, $R_M < R_C$, but the matter is still too hot to enter the magnetosphere and accrete. This regime was discussed in detail by \citet{davies79a, davies81a} and \citet{ikhsanov07a}. Following \citet{davies81a} and \citet{henrichs83a}, we can write a simple expression for the expected magnetic field given an observed spin-down rate:
\begin{equation}
    B_\text{SSP} = \sqrt{ \dot P \frac{GMI}{4\pi}}\,\text{G}
.\end{equation}

With the same spin-down rate assumptions as above, this would imply $B_\text{SSP} = 8.85\times10^{12}$\,G. While still high, this estimate is significantly lower than the ones obtain before, as expected from the different assumptions regarding the state of the source. It is difficult to estimate the luminosity at which the subsonic propeller would start, given that it strongly depends on the wind density, temperature, and turbulence just outside the magnetosphere -- which are all unknown.

\begin{table*}
  \caption{Implied magnetic field strength for three different epochs following the method given by \citet{gao21}. The values for $L$, $\dot m$, and $P$ are based on the observation at the given date (MJD). We note that for epoch 1, we do not know if the given luminosity is representative of the average luminosity during this epoch, due to the  lack of X-ray flux monitoring during that time.}
\begin{center}
\begin{tabular}{c c | c c c c c | c c c | c c c }
\hline
\hline
\\[-0.25cm]
& &  MJD & $L$\tablefootmark{a} & $\dot m$\tablefootmark{b} & $P$\tablefootmark{c} & $\dot P$\tablefootmark{d} & \multicolumn{3}{c}{low B} &   \multicolumn{3}{c}{high B} \\
&  &  &  &     &   & &  $R_M/\rsph$ & B\tablefootmark{e} & $L_\text{crit}\tablefootmark{f} $ &  $R_M/\rsph$   & B\tablefootmark{e}  & $L_\text{crit}\tablefootmark{f}$\\

\hline
\hline

\multirow{2}{*}{epoch 1} & case 1 & \multirow{2}{*}{52690.67} &  \multirow{2}{*}{6.73} & \multirow{2}{*}{64.68} & \multirow{2}{*}{1.43} & \multirow{2}{*}{-0.81}  &0.16 & 0.31 & 0.08  & 0.92 & 26.63 & 188.82\\ & case 2 &   &   &   & &   &  0.16 & 0.31 & 0.08 & 1.05 & 45.96 & 249.53 \\\hline
 \multirow{2}{*}{epoch 2} & case 1 & \multirow{2}{*}{56847.95} &  \multirow{2}{*}{9.34} & \multirow{2}{*}{75.14} & \multirow{2}{*}{1.14} & \multirow{2}{*}{-2.04}  &0.40 & 3.68 & 14.08  & 0.61 & 12.43 & 95.96\\ & case 2 &   &   &   & &   &  0.38 & 3.36 & 11.96 & 0.73 & 25.89 & 223.80 \\\hline

 \multirow{2}{*}{epoch 3}& case 1 & \multirow{2}{*}{59053.74} &  \multirow{2}{*}{5.95} & \multirow{2}{*}{61.14} & \multirow{2}{*}{1.03} & \multirow{2}{*}{-1.32}  &0.40 & 2.27 & 7.26  & 0.73 & 12.08 & 105.02\\ & case 2 &   &   &   & &   &  0.39 & 2.16 & 6.62 & 0.86 & 23.97 & 223.06 \\\hline

\end{tabular}
\tablefoot{
\tablefoottext{a}{in $10^{39}$\,\ergcms in the 0.3--10\,keV energy band}
\tablefoottext{b}{in $\mdotedd$}
\tablefoottext{c}{in seconds}
\tablefoottext{d}{in $10^{-9}$\,s\,s$^{-1}$}
\tablefoottext{e}{in  $10^{12}$\,G}
\tablefoottext{f}{in  $10^{37}$\,\ergs}

}

\label{tab:bfields}
\end{center}
\end{table*}

\subsection{The super-orbital period}

\citet{Walton16period} discovered flux variability with a period of $78.1\pm0.5$\,d in the \swift/XRT data of \ngc{}, using data from roughly weekly monitoring observations between 2014--2016. The period showed a peak-to-peak amplitude of roughly a factor of 3 and was interpreted as a super-orbital period. As can be seen in Fig.~\ref{fig_longlc}, the flux just before the off-state in 2017 follows the exact same period, with a very similar amplitude and phase. This is also true for the recovery after the off-state in 2020: the XRT flux measurements align very well within the uncertainties of the period with the extrapolated profile of the 78\,d period. Since 2020, \ngc shows a somewhat larger variability in its flux, with variations of at least a factor of 10. This increase in variability makes identifying the 78\,d period more difficult, nonetheless, the bright states in the data still largely line up  well with the peaks of the predicted profile. 
The \swift/XRT monitoring is still ongoing to investigate the stability of this period. The current data do not allow for a independent measurement of the period, so we currently cannot say if there is a change in the period after the off-state or not.

\section{Conclusions}
\label{sec_conc}

\begin{table*}
  \caption{Summary of magnetic field estimations with the different methods presented in this work.}
  \label{tab:bsummary}
\begin{center}
\begin{tabular}{l | c c  }
\hline
\hline
Method & Description & Value [$10^{13}$\,G] \\\hline\hline
\textit{Spin-Up} \\
Gao \& Li & High B & $\sim$1.2 or $\sim$2.4 \\
Gao \& Li & Low B & 0.2--0.3  \\\hline
\textit{Propeller transition} \\
Ghosh \& Lamb & $L$ of propeller at $2.2\times 10^{39}\,\ergs$ & 2.6 \\
Chaskina & advection in disk & 6 \\\hline
\textit{Spin equilibrium} \\
Chaskina & based on July 2017 data & 1.7 \\\hline
\textit{Spin-Down} \\
Parfrey & using average spin-down between 2017-2019 & 6.8 \\
Davies \& Pringle & sub-sonic propeller & 0.89 \\\hline
\end{tabular}
\end{center}
\end{table*}

We have studied the pulse period evolution of \ngc between 2003--2022. In 2017, the source entered an extended off-state, during which it dropped below the detection limit of \swift/XRT.  During this off-state the secular spin-up trend reversed, and the neutron star rotational period slowed down significantly. After the source left the off-state in mid-2020, spin-up has resumed albeit at a lower rate than before. 

We have used different methods to estimate the magnetic field of the neutron star, either based on the spin-up or spin-down strength. The main results are summarized in Table~\ref{tab:bsummary}.  In particular, we used the torque transfer model presented by \citet{gao21}  during the spin-up. We find that the calculated field strengths agree well for the two most recent spin-up episodes in 2014--2017 and 2020--2022, in particular, for the high $B$-field solution. The first epoch between 2003--2014 gives very different estimates, but due to the lack of continuous flux monitoring, the estimated X-ray flux is highly uncertain. 
The highest estimate for the magnetic field strength in our data is $\approx2.5\times10^{13}$\,G, while for the low $B$-field solution, we find values as low as $2\times10^{12}$\,G. While based on these numbers we cannot distinguish which magnetic field is present in reality, we note that for a $\sim$$10^{12}$\,G field, we would expect to see a cyclotron resonant scattering feature (CRSF) around 12\,keV, which so far has not been observed in the spectrum \citep[][]{staubert19a, Fuerst17ngc5907, Israel17}.

We also estimate the magnetic field based on the spin-down during the off-state between 2017--2019. If we assume that the source entered the propeller regime when it dropped below the detection limit of \swift/XRT at about $L=$1--2$\times10^{39}$\,\ergs, we estimate a magnetic field of $B\leq2.5\times10^{13}$\,G. Using the update disk description of \citet{Chashkina19}, we find that this luminosity for the propeller transition implies a magnetic field $B\approx1.7\times10^{13}$\,G. The spin-down torque itself  is difficult to estimate as it depends on the unknown interaction between the magnetic field and the residual matter surrounding it, be it the stellar wind or a residual accretion disk. We performed our calculations based on a description by \citet{parfrey16a} and again find that the strong spin-down can only be explained with a very high magnetic field ($\approx6.8\times10^{13}$\,G). If we assume that the source was spun down while in the subsonic propeller regime, we find a magnetic field of around $8.9\times10^{12}$\,G.

While we cannot rule out a low magnetic field directly, circumstantial evidence points clearly toward the direction of a magnetic field of a few $10^{13}$\,G in \ngc. This value is in line with previous estimates of the magnetic field \citep[e.g.,][]{Israel17,gao21} and implies that the source is accreting at very high rates.

\section*{ACKNOWLEDGEMENTS}

The authors would like to thank the referee for their useful feedback, which helped to improve the final version of the manuscript.
DJW acknowledges support from an STFC Ernest Rutherford Fellowship.
TPR acknowledges support from the Science and Technology Facilities Council (STFC) as part of the consolidated grant award ST/000244/1
This research has also made use of data obtained with \xmm, an ESA science mission
with instruments and contributions directly funded by ESA Member States, and has also
made use of public data from the \swift\ data archive. 
We thank Kyle Parfrey for the useful discussion.
This work made use of data supplied by the UK Swift Science Data Centre at the University of Leicester.


%

\label{lastpage}

\end{document}